# STUDY ON A PHASE SPACE REPRESENTATION OF QUANTUM THEORY

[1]Tokiniaina RANAIVOSON, [2]RAOELINA ANDRIAMBOLOLONA, [3]Rakotoson HANITRIARIVO, [4]Roland RABOANARY

[1] Theoretical Physics Dept, INSTN-Madagascar, *tokhiniaina@gmail.com*  
[2] Theoretical Physics Dept, INSTN-Madagascar, *raoelinasp@gmail.com, raoelinasp@yahoo.fr, instn@moov.mg*  
[3] Theoretical Physics Dept, INSTN-Madagascar, *infotsara@gmail.com, mailrivo@gmail.com*  
[4] University of Antananarivo, Faculty of Science, Department of Physics, *r_raboanary@yahoo.fr*

*Abstract* – *A study on a method for the establishment of a phase space representation of quantum theory is presented. The approach utilizes the properties of Gaussian distribution, the properties of Hermite polynomials, Fourier analysis and the current formulation of quantum mechanics which is based on the use of Hilbert space and linear operators theory. Phase space representation of quantum states and wave functions in phase space are introduced using properties of a set of functions called harmonic Gaussian functions. Then, new operators called dispersion operators are defined and identified as the operators which admit as eigenstates the basis states of the phase space representation. Generalization of the approach for multidimensional cases is shown. Examples of applications are given.*

## I. INTRODUCTION

According to the uncertainty relation, it is well known that in general the exact values of coordinate and momentum of a particle cannot be measured and known simultaneously. Let us consider a particle which move along a one dimensional axis: let $x$ be its coordinate and $p$ its momentum. Let $|\psi\rangle$ be the quantum vector state of the particle and let $\psi$ and $\tilde{\psi}$ be its wave functions respectively in coordinate and momentum representations [1]. We have

$$\psi(x) = \langle x|\psi\rangle = \frac{1}{\sqrt{2\pi\hbar}} \int \tilde{\psi}(p)\, e^{i\frac{px}{\hbar}} dp \qquad (1.1)$$

$$\tilde{\psi}(p) = \langle p|\psi\rangle = \frac{1}{\sqrt{2\pi\hbar}} \int \psi(x) e^{-i\frac{px}{\hbar}} dx \qquad (1.2)$$

$$\int |\psi(x)|^2 dx = \int |\tilde{\psi}(p)|^2 dp = 1 \qquad (1.3)$$

Let $X$ and $P$ be the mean values of $x$ and $p$

$$X = \int x|\psi(x)|^2 dx = \int \tilde{\psi}^*(p) \left[i\hbar \frac{d\tilde{\psi}(p)}{dp}\right] dp \quad (1.4)$$

$$P = \int p|\tilde{\psi}(p)|^2 dp = \int \psi^*(x)\left[-i\hbar \frac{d\psi(x)}{dx}\right] dx \quad (1.5)$$

The standard deviations $\Delta x$ and $\Delta p$ respectively of $x$ and $p$ are

$$\Delta x = \sqrt{\int (x-X)^2 |\psi(x)|^2\, dx} \qquad (1.6)$$

$$\Delta p = \sqrt{\int (p-P)^2 |\tilde{\psi}(p)|^2\, dp} \qquad (1.7)$$

Using properties of Fourier transform and Cauchy-Schwarz inequality, it may be shown that we have the inequality

$$\Delta x \Delta p \geq \frac{\hbar}{2} \qquad (1.8)$$

The inequality (1.8) is the uncertainty relation. Because of this uncertainty relation, it seems not easy to talk about phase space in quantum mechanics. However, some authors has already study the possibility of formulating quantum theory in phase space. A well known approach is based on the use of Wigner function [2],[3],[4],[5] instead of the wave function. This approach permits to obtain interesting results [6],[7],[8] but the physical interpretation of the Wigner function is not easy because it is not a positive definite distribution [9],[10].

In this work, we adopt another approach to tackle the problem of formulation of quantum theory in phase space. We utilize properties of a set of functions called harmonic Gaussian functions. Phase space states are introduced to build a basis in state space which permits to define the phase space representation. And wave functions in phase space are defined as wave functions corresponding to this representation. Then, it is shown that positive definite probability density function may be defined using these phase space wave functions. Important insights in our approach are the introduction of new operators called dispersion operators. Our approach may be applied to nonrelativistic or relativistic cases.

Statistic-probability theory and linear algebra are both used. So to avoid confusion with the use of the same words "variance" and "covariance" in statistical meaning and in linear algebra meaning (variance and covariance of a tensor), we will use the word "dispersion" and "codispersion" to designate statistical variance and covariance.

## II. HARMONIC GAUSSIAN FUNCTIONS

For positive integers $n$, let us consider the set of orthonormalized functions defined by the relations

$$\varphi_n(x, X, P, \Delta x) = \frac{H_n\left(\frac{x-X}{\sqrt{2}\Delta x}\right)}{\sqrt{2^n n! \sqrt{2\pi}\Delta x}} e^{-\left(\frac{x-X}{2\Delta x}\right)^2 + i\frac{Px}{\hbar}} \quad (2.1)$$

$$\int \varphi_n^*(x, X, P, \Delta x) \varphi_m(x, X, P, \Delta x) dx = \delta_{nm} \quad (2.2)$$

$H_n(u)$ is Hermite polynomial of degree $n$. We may establish

$$\int x|\varphi_n(x, X, P, \Delta x)|^2 dx = X \quad (2.3)$$

$$\int (x-X)^2 |\varphi_n(x, X, P, \Delta x)|^2 dx = (2n+1)(\Delta x)^2 \quad (2.4)$$

According to the relations (2.3) and (2.4), $X$ is the coordinate mean value and $(\Delta x_n)^2 = (2n+1)(\Delta x)^2$ the coordinate dispersion corresponding to $\varphi_n$. For further utilization, we call the coordinate dispersion $(\Delta x_0)^2 = (\Delta x)^2$ corresponding to $\varphi_0$ the "ground coordinate dispersion"($n=0$). As in our work [11], we call $\varphi_n$ a harmonic Gaussian function.

According to the relation (2.2), a function $\varphi_n$ may be considered as a wave function in coordinate representation, the corresponding wave function $\tilde{\varphi}_n$ in momentum representation is

$$\tilde{\varphi}_n(p, X, P, \Delta p) = \frac{1}{\sqrt{2\pi\hbar}} \int \varphi_n(x, X, P, \Delta x) e^{-i\frac{px}{\hbar}} dx$$

$$= \frac{(i)^n H_n\left(\frac{p-P}{\sqrt{2}\Delta p}\right)}{\sqrt{2^n n! \sqrt{2\pi}\Delta p}} e^{-\left(\frac{p-P}{2\Delta p}\right)^2 - ix\frac{(p-P)}{\hbar}} \quad (2.5)$$

in which

$$\Delta p = \frac{\hbar}{2\Delta x} \Leftrightarrow \Delta x \Delta p = \frac{\hbar}{2} \quad (2.6)$$

As for the case of the functions $\varphi_n$, we have the following properties for $\tilde{\varphi}_n$

$$\int \tilde{\varphi}_n^*(p, X, P, \Delta p) \tilde{\varphi}_m(p, X, P, \Delta p) dp = \delta_{nm} \quad (2.7)$$

$$\int p |\tilde{\varphi}_n(p, X, P, \Delta p)|^2 dp = P \quad (2.8)$$

$$\int (p-P)^2 |\tilde{\varphi}_n(p, X, P, \Delta p)|^2 dp = (2n+1)(\Delta p)^2 \quad (2.9)$$

According to the relation (2.8) and (2.9), $P$ is the momentum mean value and $(\Delta p_n)^2 = (2n+1)(\Delta p)^2$ the momentum dispersion corresponding to the function $\tilde{\varphi}_n$. For further utilization, we designate the momentum dispersion $(\Delta p_0)^2 = (\Delta p)^2$ corresponding to $\tilde{\varphi}_0$ the "ground momentum dispersion".

Because of the relation (2.6), it is sufficient to use only one of the parameters $\Delta x$ or $\Delta p$. So from now on we will use $\Delta p$ in all expressions. For instance, we will utilize the notation $\varphi_n(x, X, P, \Delta p)$ instead of $\varphi_n(x, X, P, \Delta x)$ for the functions $\varphi_n$.

The set of functions $\{\varphi_n\}$ and $\{\tilde{\varphi}_n\}$ are orthonormal basis in the vector space $\mathcal{L}^2$ of Lebesgue square integrable functions. Let $\psi$ and $\tilde{\psi}$ two wave functions, corresponding to a state $|\psi\rangle$, respectively in coordinate and momentum representations. The functions $\psi, \tilde{\psi}$ are elements of $\mathcal{L}^2$. We have the expansions respectively in the basis $\{\varphi_n\}$ and $\{\tilde{\varphi}_n\}$:

$$\psi(x) = \langle x|\psi\rangle = \sum_n \Psi^n(X, P, \Delta p) \varphi_n(x, X, P, \Delta p) \quad (2.10)$$

$$\tilde{\psi}(p) = \langle p|\psi\rangle = \sum_n \Psi^n(X, P, \Delta p) \tilde{\varphi}_n(p, X, P, \Delta p) \quad (2.11)$$

$$\Psi^n(X, P, \Delta p) = \int \varphi_n^*(x, X, P, \Delta p) \psi(x) dx \quad (2.12)$$

$$= \int \tilde{\varphi}_n^*(p, X, P, \Delta p) \tilde{\psi}(p) dp \quad (2.13)$$

From the relations (2.10), (2.11) and the orhonormality of the basis $\{\varphi_n\}$ and $\{\tilde{\varphi}_n\}$, we may deduce the relations

$$\int |\psi(x)|^2 dx = \int |\tilde{\psi}(p)|^2 dp = \sum_n |\Psi^n(X, P, \Delta p)|^2 \quad (2.14)$$

We may show, that between the functions $\psi, \tilde{\psi}$ and $\Psi^n$, we have also the relations

$$\psi(x) = \frac{1}{2\pi\hbar} \int \Psi^n(X, P, \Delta p) \varphi_n(x, X, P, \Delta p) dX dP \quad (2.15)$$

$$\tilde{\psi}(p) = \frac{1}{2\pi\hbar} \int \Psi^n(X, P, \Delta p) \tilde{\varphi}_n(p, X, P, \Delta p) dX dP \quad (2.16)$$

## III. PHASE SPACE REPRESENTATION

### III.1- Phase space states and phase space representation

According to the results obtained in the section II, the functions $\varphi_n(x, X, P, \Delta p)$ and $\tilde{\varphi}_n(p, X, P, \Delta p)$ may be considered as wave functions respectively in coordinate and momentum representations. If we denote $|n, X, P, \Delta p\rangle$ the state which corresponds to these wave functions, we have

$$\varphi_n(x, X, P, \Delta p) = \langle x|n, X, P, \Delta p\rangle \quad (3.1)$$
$$\tilde{\varphi}_n(p, X, P, \Delta p) = \langle p|n, X, P, \Delta p\rangle \quad (3.2)$$

According to the relations (2.3), (2.4), (2.8), (2.9), the state $|n, X, P, \Delta p\rangle$ is a state characterized by
- the coordinate mean value $X$
- the momentum mean value $P$
- the coordinate dispersion $(\Delta x_n)^2 = (2n+1)(\Delta x)^2$
- the momentum dispersion $(\Delta p_n)^2 = (2n+1)(\Delta p)^2$

a state $|n, X, P, \Delta p\rangle$ may be then considered as a state in phase space i.e. a phase space state if we define the **quantum phase space** as the set $\mathcal{P} = \{(n, X, P, \Delta p)\}$.

From the relations (2.10) we may deduce the relations

$$|\psi\rangle = \sum_n \Psi^n(X, P, \Delta P) |n, X, P, \Delta p\rangle \quad (3.3)$$

$$\Psi^n(X, P, \Delta p) = \langle n, X, P, \Delta p|\psi\rangle \quad (3.4)$$

From the relations (2.15), we may also establish the relation

$$|\psi\rangle = \frac{1}{2\pi\hbar} \int |n, X, P, \Delta p\rangle\langle n, X, P, \Delta p|\psi\rangle dX dP \quad (3.5)$$

According to the relations (3.3) and (3.5), we can deduce that the set $\{|n, X, P, \Delta p\rangle\}$ is a basis in the state space. This basis can be used to define the phase space representation. According to the relation (3.4), a function $\Psi^n(X, P, \Delta p) = \langle n, X, P, \Delta p|\psi\rangle$ is a wave function corresponding to a state $|\psi\rangle$ in this representation. The results shown in the next paragraph give more justification to this assumption.

According to the relations (3.3) and (3.5), the phase space representation that we have defined has a particularity: the expansion of a state $|\psi\rangle$ in the basis $\{|n, X, P, \Delta p\rangle\}$ may be obtained by making a summation over the index $n$ (relation(3.3)) or by making an integration on the coordinate-momentum plane $\{(X, P)\}$ (relation(3.5)). The first expansion needs the knowledge of the values of all the functions $\Psi^n(X, P, \Delta p)$ at a point $(X, P)$ for a given value of $\Delta p$. The second one needs the knowledge of the expression of one function $\Psi^n(X, P, \Delta p)$ for given values of $n$ and $\Delta p$.

### III.2- Wave functions in the phase space representation

A function $\Psi^n(X, P, \Delta p) = \langle n, X, P, \Delta p|\psi\rangle$ may be considered as a phase space wave function corresponding to a state $|\psi\rangle$. In this section, we make more justification to this assumption. Let $\rho^n, F^n$ and $G^n$ be the functions

$$\rho^n(X, P, \Delta p) = \frac{|\Psi^n(X, P, \Delta p)|^2}{2\pi\hbar} \quad (3.6)$$

$$F^n(X, \Delta p) = \int \rho^n(X, P, \Delta p)\, dP \quad (3.7)$$

$$G^n(P, \Delta p) = \int \rho^n(X, P, \Delta p)\, dX \quad (3.8)$$

Using the relations (2.12), (2.13) and properties of the functions $\varphi_n$ and $\tilde{\varphi}_n$, we may prove the relations

$$F^n(X, \Delta p) = \int |\varphi_n(x, X, P, \Delta p)|^2 |\psi(x)|^2\, dx$$

$$= \int \frac{\left|H_n\left(\frac{x-X}{\sqrt{2}\Delta x}\right)\right|^2}{2^n n! \sqrt{2\pi}\Delta x} e^{-\left(\frac{x-X}{\sqrt{2}\Delta x}\right)^2} |\psi(x)|^2\, dx \quad (3.9)$$

$$G^n(P, \Delta p) = \int_{-\infty}^{+\infty} |\tilde{\varphi}_n(p, X, P, \Delta p)|^2 |\tilde{\psi}(p)|^2\, dp \quad (3.10)$$

$$= \int \frac{\left|H_n\left(\frac{x-X}{\sqrt{2}\Delta p}\right)\right|^2}{2^n n! \sqrt{2\pi}\Delta p} e^{-\left(\frac{p-P}{\sqrt{2}\Delta p}\right)^2} |\tilde{\psi}(p)|^2\, dp \quad (3.11)$$

$$\int \rho^n(X, P, \Delta p)\, dX dP = \int F^n(X, \Delta p)\, dX = \int |\psi(x)|^2\, dx$$

$$= \int G^n(P, \Delta p)\, dP = \int |\tilde{\psi}(p)|^2\, dp = 1 \quad (3.12)$$

The analysis of these relations suggest the following interpretations

- The function $F^n$ is a representation of the probability density corresponding to the coordinate $X$ for a given value of the ground momentum dispersion $(\Delta p)^2$.

- The function $G^n$ is a representation of probability density corresponding to the momentum $P$ for a given value of the ground momentum dispersion $(\Delta p)^2$.

- The function $\rho^n$ is a representation of the probability density on the coordinate-impulsion plane $\{(X, P)\}$ for a given value of the momentum ground dispersion $(\Delta p)^2$.

According to the relation (2.14), for a given values of $X, P$ and $\Delta p$, $|\Psi^n(X, P, \Delta p)|^2$ may be also interpreted as the probability to find in a state $|n, X, P, \Delta p\rangle$ a particle which is in the state $|\psi\rangle$. These interpretations give more justification on the consideration of the functions $\Psi^n(X, P, \Delta p)$ as the wave functions in phase space representation.

### III.3-Remarkable properties of phase space states and phase space wave functions

**Property 1**: *Scalar product of two phase space states*

Let $\{|n, X, P, \Delta p\rangle\}$ and $\{|n', X', P', \Delta p\rangle\}$ two basis in the state space. From the relation (3.3), we may deduce the expression of the expansion of a state vector $|n', P', X', \Delta p\rangle$ in the basis $\{|n, X, P, \Delta p\rangle\}$

$$|n', X', P', \Delta p\rangle = \sum_n \Phi_{n'}^n(X, X', P, P', \Delta p)|n, X, \Delta x, P\rangle$$

$$\Phi_{n'}^n(X, X', P, P', \Delta p) = \langle n, X, P, \Delta p|n', X', P', \Delta p\rangle \quad (3.13)$$

$\Phi_{n'}^n(X, X', P, P', \Delta p)$ is the wave function corresponding to the state $|n', X', P', \Delta p\rangle$ in phase space representation. Let us look for the expression of this wave function

$$\Phi_{n'}^n(X, X', P, P', \Delta p) = \langle n, X, P, \Delta p|n', X', P', \Delta p\rangle$$

$$= \int \varphi_n^*(x, X, P, \Delta p) \varphi_{n'}(x, X', P', \Delta p)\, dx$$

$$= \int \tilde{\varphi}_n^*(p, X, P, \Delta p) \tilde{\varphi}_{n'}(p, X', P', \Delta p)\, dp \quad (3.14)$$

To perform the calculation of this integral, we have to use the following property of Hermite polynomials

$$\int H_n(u+a) H_{n'}(u+b) e^{-u^2}\, du$$

$$= \sum_{j=0}^{\min(n,n')} \frac{n!\, n'!\, 2^{n+n'-j}}{j!\,(n-j)!\,(n'-j)!}(a)^{n-j}(b)^{n'-j}\sqrt{\pi}$$

We find

$$\Phi_{n'}^n(X, X', P, P', \Delta p)$$

$$= R_{n'}^n(Z) e^{-\frac{1}{2}\left[\left(\frac{X-X'}{2\Delta x}\right)^2 + \left(\frac{P-P'}{2\Delta p}\right)^2 - i\left(\frac{X+X'}{2\hbar}\right)(P-P')\right]} \quad (3.15)$$

in which

$$Z = \frac{1}{2}\left(\frac{X-X'}{\sqrt{2}\Delta x} - i\frac{P-P'}{\sqrt{2}\Delta p}\right)$$

and $R_{n'}^n(Z)$ is the complex polynomial

$$R_{n'}^n(Z) = \sum_{j=0}^{\min(n,n')} \sqrt{n!\, n'!\, 2^{n+n'-2j}} \frac{(-1)^{n-j}(Z^*)^{n-j}(Z)^{n'-j}}{j!\,(n-j)!\,(n'-j)!}$$

From the relation (3.13), we may deduce the relations

$$\langle n, X, P, \Delta p | m, X', P', \Delta p \rangle = \langle m, X', P', \Delta p | n, X, P, \Delta p \rangle^*$$

$$\Rightarrow \Phi_{n'}^n(X, X', P, P', \Delta p) = [\Phi_n^{n'}(X', X, P', P, \Delta p)]^* \quad (3.16)$$

The functions $\Phi_{n'}^n(X, X', P, P', \Delta p)$ may be used to perform a basis change between $\{|n, X, P, \Delta p\rangle\}$ and $\{|n', X', P', \Delta p\rangle\}$.

**Property 2**: *Relations between two values at two points of two phase space wave functions*

Let $|\psi\rangle$ be a state, according to the relation (3.3), we can expand $|\psi\rangle$ in the basis $\{|n, X, P, \Delta p\rangle\}$ or in the basis $\{|n', X', P', \Delta p\rangle\}$.

$$|\psi\rangle = \sum_n |n, X, P, \Delta p\rangle \langle n, X, P, \Delta p|\psi\rangle$$

$$= \sum_{n'} |n', X', P', \Delta p\rangle \langle n', X', P', \Delta p|\psi\rangle$$

$$\langle n, X, P, \Delta p|\psi\rangle = \Psi^n(X, P, \Delta p)$$

$$\langle n', X', P', \Delta p|\psi\rangle = \Psi^{n'}(X', P', \Delta p)$$

Then, taking into account the relation (3.13), we may deduce the relation

$$\Psi^n(X, P, \Delta p) = \sum_{n'} \Phi_{n'}^n(X, X', P, P', \Delta p) \, \Psi^{n'}(X', P', \Delta p)$$

## IV. DISPERSION OPERATORS

For given values of the coordinate mean $X$, the momentum mean $P$ and the momentum ground dispersion $\Delta p$, The state $|n, X, P, \Delta p\rangle$, used for the definition of the phase space representation is a state characterized by the knowledge of
- the coordinate dispersion $(\Delta x_n)^2 = (2n + 1)(\Delta x)^2$
- the momentum dispersion $(\Delta p_n)^2 = (2n + 1)(\Delta p)^2$

It follows that a state $|n, X, P, \Delta p\rangle$ may be considered as an eigenstate of
- coordinate dispersion operator $\boldsymbol{\Sigma}_x$ with eigenvalue equal to $(\Delta x_n)^2 = (2n + 1)(\Delta x)^2$
- momentum dispersion operator $\boldsymbol{\Sigma}_p$ with eigenvalue equal to $(\Delta p_n)^2 = (2n + 1)(\Delta p)^2$

$$\boldsymbol{\Sigma}_x |n, X, P, \Delta p\rangle = (2n + 1)(\Delta x)^2 |n, X, P, \Delta p\rangle \quad (4.1)$$

$$\boldsymbol{\Sigma}_p |n, X, P, \Delta p\rangle = (2n + 1)(\Delta p)^2 |n, X, P, \Delta p\rangle \quad (4.2)$$

We may obtain the expression of these operators in coordinate representation by using properties of Harmonic Gaussian functions. In fact, because of the relation (3.1), in this representation we may write

$$\boldsymbol{\Sigma}_x \varphi_n(n, X, P, \Delta p) = (2n + 1)(\Delta x)^2 \varphi_n(n, X, P, \Delta p) \quad (4.3)$$

$$\boldsymbol{\Sigma}_p \varphi_n(n, X, P, \Delta p) = (2n + 1)(\Delta p)^2 \varphi_n(n, X, P, \Delta p) \quad (4.4)$$

Using directly the expressions (2.1) of Harmonic Gaussian functions, we can verify that in coordinate representation, the exact expressions of the operators $\boldsymbol{\Sigma}_x$ and $\boldsymbol{\Sigma}_p$ which satisfy exactly the relations (4.3) and (4.4) are

$$\boldsymbol{\Sigma}_x = \frac{1}{2}\left[(x - X)^2 + \frac{(\Delta x)^2}{(\Delta p)^2}\left(-i\hbar \frac{d}{dx} - P\right)^2\right] \quad (4.5)$$

$$\boldsymbol{\Sigma}_p = \frac{1}{2}\left[\frac{(\Delta p)^2}{(\Delta x)^2}(x - X)^2 + \left(-i\hbar \frac{d}{dx} - P\right)^2\right] \quad (4.6)$$

$(\Delta x)^2$ and $(\Delta p)^2$ are respectively the ground coordinate dispersion and ground momentum dispersion. They are related by the relation (2.6). $\boldsymbol{\Sigma}_x$ and $\boldsymbol{\Sigma}_p$, are related:

$$\boldsymbol{\Sigma}_p = \frac{(\Delta p)^2}{(\Delta x)^2}\boldsymbol{\Sigma}_x = \frac{\hbar^2}{4(\Delta x)^4}\boldsymbol{\Sigma}_x = \frac{4(\Delta p)^4}{\hbar^2}\boldsymbol{\Sigma}_x \quad (4.7)$$

The relations (4.5) and (4.6) suggest that if we denote $\boldsymbol{x}$ and $\boldsymbol{p}$ the operators associated respectively to the coordinate and momentum in the current formulation of quantum mechanics, we have in general (in any representation) the relation:

$$\boldsymbol{\Sigma}_x = \frac{(\Delta x)^2}{2}\left[\frac{(\boldsymbol{x} - X)^2}{(\Delta x)^2} + \frac{(\boldsymbol{p} - P)^2}{(\Delta p)^2}\right] \quad (4.8)$$

$$\boldsymbol{\Sigma}_p = \frac{(\Delta p)^2}{2}\left[\frac{(\boldsymbol{x} - X)^2}{(\Delta x)^2} + \frac{(\boldsymbol{p} - P)^2}{(\Delta p)^2}\right] \quad (4.9)$$

To find a particular expression in a given representation, we have to replace the operators $\boldsymbol{x}$ and $\boldsymbol{p}$ by their expression in the considered representation. For instance, to have the expressions in coordinate representation, we have to replace in the relations (4.8) and (4.9) $\boldsymbol{x}$ and $\boldsymbol{p}$ by

$$\boldsymbol{x} = x \qquad \boldsymbol{p} = -i\hbar \frac{d}{dx} \quad (4.10)$$

And to obtain the expressions in momentum representation, we have to perform the replacement

$$\boldsymbol{x} = i\hbar \frac{d}{dp} \qquad \boldsymbol{p} = p \quad (4.11)$$

With the mean and dispersion operators, we may also define mean quadratic operators $\overline{\boldsymbol{x^2}}$ and $\overline{\boldsymbol{p^2}}$

$$\overline{\boldsymbol{x^2}} = X^2 + \boldsymbol{\Sigma}_x \quad (4.12)$$

$$\overline{\boldsymbol{p^2}} = P^2 + \boldsymbol{\Sigma}_p \quad (4.13)$$

And we have the eigenvalue equations

$$\overline{\boldsymbol{x^2}}|n, X, P, \Delta p\rangle = [X^2 + (2n + 1)(\Delta x)^2]|n, X, P, \Delta p\rangle \quad (4.14)$$

$$\overline{\boldsymbol{p^2}}|n, X, P, \Delta p\rangle = [P^2 + (2n + 1)(\Delta p)^2]|n, X, P, \Delta p\rangle \quad (4.15)$$

## V. MULTIDIMENSIONAL GENERALIZATION

We can generalize the results obtain in the previous sections to the cases of higher dimension than one. We study the case of uncorrelated and correlated variables one after the others. Calculation related to linear algebra are based on formulation given in [12]

## V.1- Case of uncorrelated variables

Let us consider a quantum system which may be described with a position vector $\vec{x}$ belonging to a $N-$dimensional vector space $E_N$, $\vec{x} = x^l \vec{e}_l$ ($l = 1,2, \ldots N$). $\{\vec{e}_l\}$ is a basis in the space $E_N$. We introduce the momentum associated to $\vec{x}$ as a covector on $E_N$ that we denote $\breve{p}$, $\breve{p} = p_l \breve{e}^l$ in which $\{\breve{e}^l\}$ is the cobasis of the basis $\{\vec{e}_l\}$ i.e. $\{\breve{e}^l\}$ is the basis in the dual $E_N^*$ of $E_N$ verifying $\breve{e}^l(\vec{e}_j) = \delta_j^l$ [12]. Let $|\vec{x}\rangle$ and $|\breve{p}\rangle$ be the basis states for the coordinate and momentum representations. Any state $|\psi\rangle$ of the system may be expanded in the basis $\{|\vec{x}\rangle\}$ and $\{|\breve{p}\rangle\}$

$$|\psi\rangle = \int |\vec{x}\rangle \langle \vec{x}|\psi\rangle d^N x = \int |\breve{p}\rangle \langle \breve{p}|\psi\rangle d^N p \quad (5.1)$$

$$\langle \vec{x}|\psi\rangle = \psi(\vec{x}) = \frac{1}{(\sqrt{2\pi\hbar})^N} \int \tilde{\psi}(\breve{p}) e^{i\frac{p_l x^l}{\hbar}} d^N p \quad (5.2)$$

$$\langle \breve{p}|\psi\rangle = \tilde{\psi}(\breve{p}) = \frac{1}{(\sqrt{2\pi\hbar})^N} \int \psi(\vec{x}) e^{-i\frac{p_l x^l}{\hbar}} d^N x \quad (5.3)$$

$\psi$ and $\tilde{\psi}$ are respectively the wave functions in coordinate and momentum representations. The variables are uncorrelated if we have the relations.

$$|\vec{x}\rangle = \bigotimes_{l=1}^{N} |x^l\rangle = |x^1\rangle \otimes |x^2\rangle \ldots \otimes |x^N\rangle \quad (5.4)$$

$$|\breve{p}\rangle = \bigotimes_{l=1}^{N} |p_l\rangle = |p_1\rangle \otimes |p_2\rangle \ldots \otimes |p_N\rangle \quad (5.5)$$

$$|\psi\rangle = \bigotimes_{l=1}^{N} |\psi^l\rangle = |\psi^1\rangle \otimes |\psi^2\rangle \ldots \otimes |\psi^N\rangle \quad (5.6)$$

$$\psi(\vec{x}) = \prod_{l=1}^{N} \psi^l(x^l) = \psi^1(x^1)\psi^2(x^2)\ldots\psi^N(x^N) \quad (5.7)$$

$$\tilde{\psi}(\breve{p}) = \prod_{l=1}^{N} \tilde{\psi}^l(p_l) = \tilde{\psi}^1(p_1)\tilde{\psi}^2(p_2)\ldots\tilde{\psi}^N(p_N) \quad (5.8)$$

$$\psi^l(x^l) = \langle x^l|\psi^l\rangle \quad \tilde{\psi}^l(p_l) = \langle p_l|\psi^l\rangle \quad (5.9)$$

$|x^l\rangle$ is a one dimensional coordinate state and $|p_l\rangle$ is one dimensional momentum state. $\psi^l$ and $\tilde{\psi}^l$ are the wave functions corresponding to the variables $x^l$ and $p_l$ respectively in coordinate and momentum representations.

We may introduce the phase space state $|n^l, X^l, P_l, \Delta p_l\rangle$ corresponding to the variable $x^l$ and phase space wave functions $\Psi^{n^l}$ such as

$$\langle x^l|n^l, X^l, P_l, \Delta p_l\rangle = \varphi_{n^l}(x^l, X^l, P_l, \Delta p_l) \quad (5.10)$$

$$\langle p_l|n^l, X^l, P_l, \Delta p_l\rangle = \tilde{\varphi}_{n^l}(p_l, X^l, P_l, \Delta p_l) \quad (5.11)$$

$$\Psi^{n^l}(X^l, P_l, \Delta p_l) = \langle \psi^l|n^l, X^l, P_l, \Delta p_l\rangle$$
$$= \int \varphi_{n^l}^*(x^l, X^l, P_l, \Delta p_l)\, \psi^l(x^l) dx^l$$
$$= \int \tilde{\varphi}_{n^l}^*(p_l, X^l, P_l, \Delta p_l)\, \tilde{\psi}^l(p_l) dp_l$$

$$\psi^l(x^l) = \sum_{n^l} \Psi^{n^l}(X^l, P_l, \Delta p_l)\varphi_{n^l}(x^l, X^l, P_l, \Delta p_l)$$
$$= \frac{1}{2\pi\hbar} \int \Psi^l(X^l, P_l, \Delta p_l)\varphi_{n^l}(x^l, X^l, P_l, \Delta p_l) dX^l dP_l$$

We may define the state

$$|n, \vec{X}, \breve{P}, [\Delta p]\rangle = \bigotimes_{l=1}^{N} |n^l, X^l, P_l, \Delta p_l\rangle \quad (5.12)$$

We have for the wave functions

$$\Psi^n(\vec{X}, \breve{P}, [\Delta p]) = \langle \psi|n, \vec{X}, \breve{P}, [\Delta p]\rangle = \prod_{l=1}^{N} \Psi^{n^l}(X^l, P_l, \Delta p_l)$$
$$= \prod_{l=1}^{N} \int \varphi_{n^l}^*(x^l, X^l, P_l, \Delta p_l)\, \psi^l(x^l) dx^l$$
$$= \int \varphi_n^{N*}(\vec{x}, \vec{X}, \breve{P}, [\Delta p])\, \psi(\vec{x}) d^N x \quad (5.13)$$

$$\psi(\vec{x}) = \prod_{l=1}^{N} \sum_{n^l} \Psi^{n^l}(X^l, P_l, \Delta p_l)\, \varphi_{n^l}(x^l, X^l, P_l, \Delta p_l)$$
$$= \sum_{(n,N)} \Psi^n(\vec{X}, \breve{P}, [\Delta p])\, \varphi_n^N(\vec{x}, \vec{X}, \breve{P}, [\Delta p]) \quad (5.14)$$
$$= \prod_{l=1}^{N} [\frac{1}{2\pi\hbar} \int \Psi^{n^l}(X^l, P_l, \Delta p_l)\varphi_{n^l}(x^l, X^l, P_l, \Delta p_l) dX^l dP_l]$$
$$= \frac{1}{(2\pi\hbar)^N} \int \Psi^n(\vec{X}, \breve{P}, [\Delta p])\, \varphi_n^N(\vec{x}, \vec{X}, \breve{P}, [\Delta p]) d^N X d^N P \quad (5.15)$$

In these relations, $n$ is the $N$-uplet $n = (n^1, \ldots, n^N) \in \mathbb{N}^N$. The summation in (5.14) is to be performed for all possible values of $N$-uplet $n$. $\vec{X} = X^l \vec{e}_l$, $\breve{P} = P_l \breve{e}^l$. $[\Delta p]$ is the diagonal matrix $[\Delta p] = Diag(\Delta p_1, \ldots, \Delta p_N)$. $\varphi_n^N$ is the function $\varphi_n^N(\vec{x}, \vec{X}, \breve{P}, [\Delta p]) = \langle \vec{x}|n, \vec{X}, \breve{P}, [\Delta p]\rangle$

$$\varphi_n^N(\vec{x}, \vec{X}, \breve{P}, [\Delta p]) = \prod_{\mu=1}^{N} \varphi_{n^l}(x^l, X^l, P_l, \Delta p_l) \quad (5.16)$$

We may call $\varphi_n^N$ an uncorrelated multidimensional harmonic Gaussian functions. We may introduce the matrix

$$[\Delta x] = Diag(\Delta x^1, \ldots, \Delta x^N)$$
$$= Diag\left(\frac{\hbar}{2\Delta p_1}, \frac{\hbar}{2\Delta p}, \ldots, \frac{\hbar}{2\Delta p_N}\right) = \frac{\hbar}{2}[\Delta p]^{-1} \quad (5.17)$$

Then with the notations

$$|n| = n^1 + n^2 + \cdots n^N \quad n! = n^1! \, n^2! \ldots n^N!$$

$$\det[\Delta x] = \prod_{l=1}^{N} \Delta x^l \quad H_n(\vec{x}, \vec{X}, [\Delta p]) = \prod_{l=1}^{N} H_{n^l}(\frac{x^l - X^l}{\sqrt{2}\Delta x^l})$$

$$\left(\frac{\vec{x} - \vec{X}}{2[\Delta x]}\right)^2 = \sum_{l=1}^{N} \left(\frac{x^l - X^l}{2\Delta x^l}\right)^2$$

$$\breve{P}\vec{x} = P_l x^l = P_1 x^1 + \cdots + P_N x^N$$

the expression of $\varphi_n^N$ takes the form

$$\varphi_n^N(\vec{x}, \vec{X}, \breve{P}, [\Delta p]) = \frac{H_n(\vec{x}, \vec{X}, [\Delta p]) e^{-(\frac{\vec{x}-\vec{X}}{2[\Delta x]})^2 + i\breve{P}\vec{x}}}{\sqrt{2^{|n|} n! (\sqrt{2\pi})^N det[\Delta x]}} \quad (5.18)$$

From the properties (2.2), (2.3) and (2.4) of the one dimensional harmonic function, we may deduce for the $\varphi_n^N$

$$\int \varphi_n^{N*}(\vec{x}, \vec{X}, \breve{P}, [\Delta p]) \varphi_m^N(\vec{x}, \vec{X}, \breve{P}, [\Delta p]) d^N x = \delta_{nm} \quad (5.19)$$

$$\int x^l |\varphi_n^N(\vec{x}, \vec{X}, \breve{P}, [\Delta p])|^2 d^N x = X^l \quad (5.20)$$

$$\int (x^l - X^l)(x^l - X^l) |\varphi_n^N(\vec{x}, \vec{X}, \breve{P}, [\Delta p])|^2 d^N x$$
$$= (2n^l + 1)(\Delta x^l)^2 \delta^{lj} \quad (5.21)$$

From the relations (2.7), (2.8) and (2.9), we may establish analogous properties for the functions

$$\tilde{\varphi}_n^N(\breve{p}, \vec{X}, \breve{P}, [\Delta p]) = \langle \breve{p} | n, \vec{X}, \breve{P}, [\Delta p] \rangle$$

Then we call $\vec{X} = X^l \vec{e}_l$ the position mean vector and the diagonal matrix

$$[\Delta x]_n = Diag[(2n^1 + 1)\Delta x^1, \ldots, (2n^N + 1)\Delta x^N)]$$

the coordinate dispersion matrix corresponding to $\varphi_n^N$. We call the dispersion matrix $[\Delta x]_0$ corresponding to $\varphi_0^N$, i.e. for $n = (0,0,\ldots 0)$, the ground position dispersion matrix.
We call $\breve{P} = p_l \breve{e}^l$ the momentum mean covector and the matrix

$$[\Delta p]_n = Diag[(2n^1 + 1)\Delta p_1, \ldots, (2n^N + 1)\Delta p_N]$$

the momentum dispersion matrix corresponding to $\tilde{\varphi}_n^N$. We call the momentum dispersion matrix $[\Delta p]_0$ corresponding to $\tilde{\varphi}_n^N$ the ground momentum dispersion matrix.

From the result obtained for one dimensional case in the section 4, we may deduce that for uncorrelated variables $x^l$, coordinates and momentum dispersion operators may be associated to each doublet $(x^l, p_l)$. According to the relations (4.8) and (4.9), if we denote $\boldsymbol{x}^l$ and $\boldsymbol{p}_l$ the operators associated respectively to $x^l$ and $p_l$ in current formulation of quantum mechanics, we may associate to them the dispersion operators $\boldsymbol{\Sigma}_{\vec{x}}^{ll}$ and $\boldsymbol{\Sigma}_{ll}^{\breve{p}}$ such as

$$\boldsymbol{\Sigma}_{\vec{x}}^{ll} = \frac{(\Delta x^l)^2}{2} \left[\frac{(\boldsymbol{x}^l - X^l)^2}{(\Delta x^l)^2} + \frac{(\boldsymbol{p}_l - P_l)^2}{(\Delta p_l)^2}\right] \quad (5.22)$$

$$\boldsymbol{\Sigma}_{ll}^{\breve{p}} = \frac{(\Delta p_l)^2}{2} \left[\frac{(\boldsymbol{x}^l - X^l)^2}{(\Delta x^l)^2} + \frac{(\boldsymbol{p}_l - P_l)^2}{(\Delta p_l)^2}\right] \quad (5.23)$$

The eigenstates of these operators are the states in the relations (5.12), the eigenvalue equations are

$$\boldsymbol{\Sigma}_{\vec{x}}^{ll} |n, \vec{X}, \breve{P}, [\Delta p]\rangle = (2n^l + 1)(\Delta x^l)^2 |n, \vec{X}, \breve{P}, [\Delta p]\rangle$$

$$\boldsymbol{\Sigma}_{ll}^{\breve{p}} |n, \vec{X}, \breve{P}, [\Delta p]\rangle = (2n^l + 1)(\Delta p_l)^2 |n, \vec{X}, \breve{P}, [\Delta p]\rangle$$

we may also define quadratic mean operators

$$\overline{(x^l)^2} = (X^l)^2 + \boldsymbol{\Sigma}_{\vec{x}}^{ll} \quad (5.24)$$

$$\overline{(p_l)^2} = (P_l)^2 + \boldsymbol{\Sigma}_{ll}^{\breve{p}} \quad (5.25)$$

### V.2 Case of correlated variables

To study correlated variables, we introduce a more general definition of multidimensional harmonic Gaussian functions with correlated variables. To obtain correlated variables, we consider a linear transformation in the space $E_N$ which mix the uncorrelated variables $x^l$. Let $\Lambda$ be a linear transformation in the space $E_N$ and let $\Lambda_j^l$ be its matrix elements in the basis $\{\vec{e}_l\}$. We introduce a new position vector $\vec{y} = y^l \vec{e}_l$ with its mean $\vec{Y} = Y^l \vec{e}_l$ and the correspondent momentum covector $\breve{k} = k_l \breve{e}^l$ with its mean $\breve{K} = K_l \breve{e}^l$ by the relation

$$\begin{cases} y^l = \Lambda_j^l x^j \\ Y^l = \Lambda_j^l X^j \\ k_l = V_l^j p_j \\ K_l = V_l^j P_j \end{cases} \Leftrightarrow \begin{cases} x^j = V_l^j y^l \\ X^j = V_l^j Y^l \\ p_j = \Lambda_j^l k_l \\ P_j = \Lambda_j^l K_l \end{cases} \quad (5.26)$$

In the relation (5.26), we have introduced the inverse $V = \Lambda^{-1}$ of $\Lambda$ and $V_j^l$ are the elements of the matrix $V$.
We have

$$\breve{K}.\vec{y} = K_l y^l = \Lambda_j^l P_l V_q^j x^q = P_l \delta_q^l x^q = P_l x^l = \breve{P}.\vec{x} \quad (5.27)$$

$$\sum_{r=1}^{N} \left(\frac{x^r - X^r}{2\Delta x^r}\right)^2 = \sum_{r=1}^{N} \frac{V_l^r V_j^r}{(2\Delta x^r)^2} (y^l - Y^l)(y^j - Y^j) \quad (5.28)$$

$$\sum_{r=1}^{N} \left(\frac{p_r - P_r}{2\Delta p_r}\right)^2 = \sum_{r=1}^{N} \frac{\Lambda_r^l \Lambda_r^j}{(2\Delta k_r)^2} (k_l - K_l)(k_j - K_j) \quad (5.29)$$

Let us introduce the two order contravariant tensor $\mathcal{A} = \mathcal{A}^{lj} \vec{e}_l \otimes \vec{e}_j$ and a two order covariant tensor $\mathcal{B} = \mathcal{B}_{lj} \breve{e}^l \otimes \breve{e}^j$ such as

$$\mathcal{A}^{lj} = \sum_{r=1}^{N} \frac{\hbar^2 \Lambda_r^l \Lambda_r^j}{(2\Delta p_r)^2} = \Lambda_r^l \Lambda_r^j \Delta x^r \Delta x^r \quad (5.30)$$

$$\mathcal{B}_{lj} = \sum_{r=1}^{N} \frac{\hbar^2 V_l^r V_j^r}{(2\Delta x^r)^2} = V_l^r V_j^r \Delta p_r \Delta p_r \quad (5.31)$$

There is a summation over the index $r$ in the last term of these equalities. The inverses of the relations (5.30) and (5.31) are

$$(\Delta x^r)^2 = V_l^r V_j^r \mathcal{A}^{lj} \qquad (\Delta p_r)^2 = \Lambda_r^l \Lambda_r^j \mathcal{B}_{\alpha\beta} \quad (5.32)$$

We may also establish the relations

$$\mathcal{A}^{lr} \mathcal{B}_{rj} = \frac{\hbar^2}{4} \delta_j^l \qquad det\mathcal{A} = \frac{\hbar^2}{4det\mathcal{B}} \quad (5.33)$$

If we choose the linear transformation $\Lambda$ such as $(det\Lambda)^2 = 1$, we may establish from the relation (5.32)

$$det[\Delta x] = \sqrt{det\mathcal{A}} \quad det[\Delta p] = \sqrt{det\mathcal{B}} \quad (5.34)$$

Introducing the expressions of $\mathcal{A}^{lj}$ and $\mathcal{B}_{lj}$ in the relations (5.28) and (5.29), we may deduce

$$\sum_{r=1}^{N} (\frac{x^r - X^r}{2\Delta x^r})^2 = \frac{\mathcal{B}_{lj}}{\hbar^2}(y^l - Y^l)(y^j - Y^j) \quad (5.35)$$

$$\sum_{r=1}^{N} (\frac{p_r - P_r}{2\Delta p_r})^2 = \frac{\mathcal{A}^{lj}}{\hbar^2}(k_l - K_l)(k_j - K_j) \quad (5.36)$$

As the tensors $\mathcal{B}$ and $\mathcal{A}$ are by definitions bilinear forms respectively on $E_N$ and on $E_N^*$ [12], we may write the relations (5.35) and (5.36) in more condensed forms

$$\sum_{r=1}^{N} (\frac{x^r - X^r}{2\Delta x^r})^2 = \frac{1}{\hbar^2}\mathcal{B}(\vec{y} - \vec{Y})(\vec{y} - \vec{Y}) \quad (5.37)$$

$$\sum_{r=1}^{N} (\frac{p_r - P_r}{2\Delta p_r})^2 = \frac{1}{\hbar^2}\mathcal{A}(\bar{k} - \bar{K})(\bar{k} - \bar{K}) \quad (5.38)$$

If we define the functions $\mathcal{H}_n(\vec{y}, \vec{Y}, \mathcal{B}) = H_n(\vec{x}, \vec{X}, [\Delta p])$, we can deduce from the expression (5.18) of $\varphi_n^N$ the definition of the wave functions $\chi_n^N$ associated to the correlated variables $y^\mu$ in coordinate representation

$$\chi_n^N(\vec{y}, \vec{Y}, \overline{K}, \mathcal{B}) = \varphi_n^N(\vec{x}, \vec{X}, \bar{P}, [\Delta p])$$

$$= \frac{\mathcal{H}_n(\vec{y}, \vec{Y}, \mathcal{B})e^{-\frac{1}{\hbar^2}\mathcal{B}(\vec{y}-\vec{Y})(\vec{y}-\vec{Y})+i\overline{K}\vec{y}}}{\sqrt{2^{|n|}n!(\sqrt{2\pi})^N\sqrt{det\mathcal{A}}}} \quad (5.39)$$

By analogy, we may define the wave function in momentum representation $\tilde{\chi}_n^N(\vec{y}, \vec{Y}, \overline{K}, \mathcal{B}) = \tilde{\varphi}_n^N(\vec{p}, \vec{X}, \bar{P}, [\Delta p])$. Then, we may introduce the phase space state $|n, \vec{Y}, \overline{K}, \mathcal{B}\rangle$ such as

$$\chi_n^N(\vec{y}, \vec{Y}, \overline{K}, \mathcal{B}) = \langle \vec{y}|n, \vec{Y}, \overline{K}, \mathcal{B}\rangle \quad (5.40)$$

$$\tilde{\chi}_n^N(\vec{y}, \vec{Y}, \overline{K}, \mathcal{B}) = \langle \bar{k}|n, \vec{Y}, \overline{K}, \mathcal{B}\rangle \quad (5.41)$$

The relation (5.40) and (5.41) may be used to describe basis change from the basis $\{|\vec{y}\rangle\}$ and $\{|\bar{k}\rangle\}$ to the basis $\{|n, \vec{Y}, \overline{K}, \mathcal{B}\rangle\}$. This basis change corresponds to the change from coordinate and momentum representations to the phase space representation and then defines this latest one.

Using the relations (5.26), we may find the expressions of dispersion operators for the case of correlated variables $y^l$.

$$\Sigma_{\vec{y}}^{lj} = \Lambda_r^l \Lambda_r^j \Sigma_{\vec{x}}^{rr} \quad \Sigma_{lj}^{\bar{k}} = V_l^r V_j^r \Sigma_{rr}^{\bar{p}} \quad (5.42)$$

Using the relations (5.17), (5.30) and (5.31), we may establish

$$\Sigma_{\vec{y}}^{lj} = \frac{1}{2}(y^l - Y^l)(y^j - Y^j)$$

$$+ \frac{2}{\hbar^2}\mathcal{A}^{lr}\mathcal{A}^{jq}(k_r - K_r)(k_q - K_q) \quad (5.43)$$

$$\Sigma_{lj}^{\bar{k}} = \frac{1}{2}(k_l - K_l)(k_j - K_j)$$

$$+ \frac{2}{\hbar^2}\mathcal{B}_{lr}\mathcal{B}_{jq}(y^r - Y^r)(y^q - Y^q) \quad (5.44)$$

We have the eigenvalues equations

$$\Sigma_{\vec{y}}^{lj}|n, \vec{Y}, \overline{K}, \mathcal{B}\rangle = [\sum_{r=1}^{N}(\Lambda_r^l \Lambda_r^j)(2n^r + 1)](\Delta x^r)^2]|n, \vec{Y}, \overline{K}, \mathcal{B}\rangle$$

$$\Sigma_{lj}^{\bar{k}}|n, \vec{Y}, \overline{K}, \mathcal{B}\rangle = [\sum_{\rho=1}^{N}V_l^r V_j^r [(2n^r + 1)(\Delta p_r)^2]|n, \vec{Y}, \overline{K}, \mathcal{B}\rangle$$

the states $|n, \vec{Y}, \overline{K}, \mathcal{B}\rangle$ are eigenstates of the operators $\Sigma_{\vec{y}}^{lj}$ and $\Sigma_{lj}^{\bar{k}}$ respectively with the eigenvalues.

$$\mathcal{A}_n^{lj} = \sum_{r=1}^{N}(2n^r + 1)\Lambda_r^l \Lambda_r^j (\Delta x^r)^2 \quad (5.45)$$

$$\mathcal{B}_{lj}^n = \sum_{r=1}^{N}(2n^r + 1)V_l^r V_j^r (\Delta p_r)^2 \quad (5.46)$$

For the case $n = (0, 0, \ldots 0)$, by taking into account the relation (5.38) and (5.39), we may obtained

$$\mathcal{A}_0^{lj} = \sum_{r=1}^{N}\Lambda_r^l \Lambda_r^j (\Delta x^r)^2 = \mathcal{A}^{lj} \quad (5.47)$$

$$\mathcal{B}_{lj}^0 = \sum_{r=1}^{N}V_l^r V_j^r (\Delta p_r)^2 = \mathcal{B}_{lj} \quad (5.48)$$

## VI. EXAMPLE OF APPLICATION IN NONRELATIVISTIC QUANTUM MECHANICS

In nonrelativistic mechanics, the classical expression of the energy of a "free particle" with mass $m$ may be written as

$$E = \frac{\check{p}^2}{2m} = \frac{(p_1)^2 + (p_2)^2 + (p_3)^2}{2m} \quad (6.1)$$

in which $\check{p} = p_l \check{e}^l$ is the momentum covector. If we denote $\vec{x} = x^l \vec{e}_l$ ($l = 1,2,3$) the position vector of the particle in the three dimensional Euclidian space and $t$ the time, the elementary equation of motion of the particle is

$$x^l(t) = \frac{p_l}{m}t + x^l(0) \quad (6.2)$$

If we consider quantum mechanics, the wave function of the free particle, of momentum $\check{p}$ and energy $E = (\check{p})^2/2m$, respectively in coordinate and momentum representation are considered as the functions

$$\psi_E(\vec{x}) = \frac{1}{(\sqrt{2\pi\hbar})^3} e^{-i\frac{\tilde{p}\cdot\vec{x}}{\hbar}} \quad (6.3a)$$

$$\tilde{\psi}_E(\tilde{p}') = \delta^{(3)}(\tilde{p}' - \tilde{p}) \quad (6.3b)$$

But these functions don't fulfill the normalization relation

$$\int |\psi(\vec{x})|^2 \, d^3x = \int |\tilde{\psi}(\tilde{p}')|^2 \, d^3p' = 1 \quad (6.4)$$

We remark that this difficulty is a consequence of the fact that the limit $\Delta p_l \to 0$ is intrinsically assumed. This difficulty may be solved by introducing our approach in which this limit is not assumed even for a "free particle".

For the application of our approach, we suppose that the variables are uncorrelated and we assume the following hypothesis

**Hypothesis 1**

The mean values $X^l$ and $P_l$ associated to each variables $x^l$ and $p_l$ are identified to the "classical values" of these quantities. So we have a "mean trajectory" of the particle defined by the equation

$$X^l = X^l(t) = \frac{P_l}{m} t + X^l(0) \quad (6.5)$$

**Hypothesis 2**

To the square of the momentum is associated the quadratic mean operator $\overline{\tilde{p}^2}$ :

$$\overline{\tilde{p}^2} = \overline{(p_1)^2} + \overline{(p_2)^2} + \overline{(p_3)^2} \quad (6.6)$$

in which the expression of a quadratic mean operator $\overline{(p_l)^2}$ may be deduced from the general expression (5.33)

$$\overline{(p_l)^2} = (P_l)^2 + \Sigma_{ll}^{\tilde{p}} \quad (6.7)$$

The expression of the momentum dispersion operator $\Sigma_{ll}^{\tilde{p}}$ may be deduced from the general expression (5.29)

$$\Sigma_{ll}^{\tilde{p}} = \frac{(\Delta p_l)^2}{2} \left[ \frac{(x^l - X^l)^2}{(\Delta x^l)^2} + \frac{(p_l - P_l)^2}{(\Delta p_l)^2} \right] \quad (6.8)$$

the operators $x^l$ and $p_l$ are the operators associated to the quantities $x^l$ and $p_l$ in ordinary quantum mechanics. We have for instance in coordinate representation

$$x^l = x^l \qquad p_l = -i\hbar \frac{\partial}{\partial x^l} \quad (6.9)$$

**Hypothesis 3**

There is an Hamiltonian operator $H$ which admits as eigenvalues the values of the energy $E$. The expression of the Hamiltonian $H$ may be deduced from the classical expression of the energy by replacing $\tilde{p}^2$ by the quadratic mean operator $\overline{\tilde{p}^2}$. According to this hypothesis, we have for the free particle the expression of the Hamiltonian operator

$$H = \frac{\overline{\tilde{p}^2}}{2m} = \sum_{l=1}^{3} \frac{(P_l)^2}{2m} + \frac{\Sigma_{ll}^{\tilde{p}}}{2m} \quad (6.10)$$

From the relation (6.10), we may deduce that the eigenstates $|\psi\rangle$ of the Hamiltonian are the eigenstates of the momentum dispersion operators.

$$H|n, \vec{X}, \vec{P}, [\Delta p]\rangle = [\frac{(P_l)^2}{2m} + \frac{\Sigma_{ll}^{\tilde{p}}}{2m}]|n, \vec{X}, \vec{P}, [\Delta p]\rangle$$

$$= \frac{(P_l)^2}{2m} + (2n^l + 1)\frac{(\Delta p_l)^2}{2m}|n, \vec{X}, \vec{P}, [\Delta p]\rangle$$

the corresponding eigenvalues which are identified to the possible values of the energy are

$$E_n = \sum_{l=1}^{3} \frac{(P_l)^2}{2m} + (2n^l + 1)\frac{(\Delta p_l)^2}{2m} \quad (6.11)$$

$n$ is the triplet $n = (n^1, n^2, n^3)$,. $\Delta p_l$ are the elements of the diagonal matrix $[\Delta p]_n = Diag[\Delta p_1, \Delta p_2, \Delta p_3]$ which is the ground momentum dispersion matrix.

According to these results, the eigenstates of the Hamiltonian operators are the phase space states $|\psi_n\rangle = |n, \vec{X}, \vec{P}, [\Delta p]\rangle$. The corresponding wave functions, $\psi_n$ and $\tilde{\psi}_n$, respectively in coordinate and momentum representations are

$$\psi_n(\vec{x}) = \langle \vec{x}|\psi_n\rangle = \langle \vec{x}|n, \vec{X}, \vec{P}, [\Delta p]\rangle = \varphi_n^3(\vec{x}, \vec{X}, \vec{P}, [\Delta p])$$

$$= \prod_{l=1}^{3} \varphi_{n^l}(x^l, X^l, P_l, \Delta p_l) \quad (6.12)$$

$\varphi_{n^l}$ is a one dimensional harmonic Gaussian function.

$$\tilde{\psi}_n(\tilde{p}) = \langle \tilde{p}|\psi_n\rangle = \langle \tilde{p}|n, \vec{X}, \vec{P}, [\Delta p]\rangle = \varphi_n^3(\tilde{p}, \vec{X}, \vec{P}, [\Delta p])$$

$$= \prod_{l=1}^{3} \tilde{\varphi}_{n^l}(p_l, X^l, P_l, \Delta p_l) \quad (6.13)$$

in which

$$\tilde{\varphi}_{n^l}(p_l, X^l, P_l, \Delta p_l) = \frac{1}{\sqrt{2\pi\hbar}} \int \varphi_{n^l}(x^l, X^l, P_l, \Delta p_l) \, dx^l$$

Unlike the functions in the relations $(6.3a)$ and $(6.3b)$, the functions in (6.12) and (6.13) fulfill the normalization relation (6.4).

The wave functions in phase space representation are

$$\Psi_n^{n'}(\vec{X}', \vec{X}, \vec{P}', \vec{P}, [\Delta p]) = \langle n', \vec{X}', \vec{P}'|\psi_n\rangle$$

$$= \langle n', \vec{X}', \vec{P}', [\Delta p]|n, \vec{X}, \vec{P}, [\Delta p]\rangle$$

$$= \prod_{l=1}^{3} \Phi_{n^l}^{n'^l}(X'^l, X^l, P'^l, P^l, \Delta p_l)$$

The expression of a function $\Phi_{n^l}^{n'^l}$ can be deduced easily from the relation (3.15).

The relation (6.11) shows that the energy of the free particle is equal to the sum of a classical kinetic term and a "quantum term" which is a linear function of the square of the momentum ground dispersions $(\Delta p_l)^2$.

According to the hypothesis 1, the particle has a "mean trajectory". The equation of this trajectory is given by the relation (6.5). The first term in the expression of the energy which is equal to the classical kinetic energy may be associated with the "mean motion" corresponding to the "mean trajectory". And the second term may be associated to the "quantum effect" which results from the dispersion of the values of momentum and coordinates around their mean values.

These results obtained for the case of a "free particle" may be generalized: a general system may have a "mean trajectory" which is a classical trajectory in phase space and quantum effect appear in the dispersion of the values of coordinates and momentum around this "mean trajectory": In our method, these facts are described by the introduction of the dispersion operators.

The study about the relation between uncorrelated and correlated variables that we have considered give a possibility to include in the analysis the study of linear change in the coordinate, for instance a rotation of the coordinates axis.

## VII. EXAMPLE OF APPLICATION IN RELATIVISTIC QUANTUM THEORY

It is possible to utilize our approach in the case of relativistic theories. As an example, we show in this section an application in the establishment of field equation in relativistic field theory. We consider the case of scalar field.

### *VII.1 Recall about the Klein-Gordon equation*

We consider the Minkowski space with signature $(+,-,-,-)$. Let $\begin{bmatrix} x^0 \\ \vec{x} \end{bmatrix}$ and $[p_0 \quad \vec{p}]$ the position four-vector and momentum four-covector. Let $x^l$ ($l = 1,2,3$) and $p_l$ be respectively the components of the vector $\vec{x}$ and the covector $\vec{p}$. In the theory of special relativity, the relation between the components of the four-momentum is

$$p_\mu p^\mu = m^2 c^2 \Leftrightarrow g^{\mu\nu} p_\mu p_\nu - m^2 c^2 = 0 \quad (7.1)$$

If we make the replacement $p_\mu \to i\hbar\partial_\mu$ in this relation we can deduce the operatorial relation

$$g^{\mu\nu}\partial_\mu\partial_\nu + \frac{m^2 c^2}{\hbar^2} = 0 \quad (7.2)$$

From this relation we can deduce the Klein-Gordon equation

$$(g^{\mu\nu}\partial_\mu\partial_\nu + \frac{m^2 c^2}{\hbar^2})\phi = 0 \quad (7.3)$$

In relativistic field theory, the function $\phi$ wich fulfills this equation is a scalar field. Now, we expect to obtain a new equation for scalar field by using our approach. We have to introduce quadridimensional harmonic Gaussian functions.

### *VII.2 Quadridimensional harmonic Gaussian functions*

If we suppose that the components $x^\mu$ ($\mu = 0,1,2,3$) of the four vector position are uncorrelated, we may define, according to the results in the section V, the uncorrelated quadridimensional harmonic Gaussian function

$$\varphi_n^4(x^0, \vec{x}, X^0, \vec{X}, P_0, \vec{P}, [\Delta p]) = \prod_{\mu=0}^{3} \varphi_{n^\mu}(x^\mu, X^\mu, P_\mu, \Delta p_\mu)$$

These functions are eigenfunctions of the dispersion operators

$$\Sigma_x^{\mu\mu} = \frac{(\Delta x^\mu)^2}{2}\left[\frac{(x^\mu - X^\mu)^2}{(\Delta x^\mu)^2} + \frac{(i\hbar\partial_\mu - P_\mu)^2}{(\Delta p_\mu)^2}\right] \quad (7.4)$$

$$\Sigma_{\mu\mu}^p = \frac{(\Delta p_\mu)^2}{2}\left[\frac{(x^\mu - X^\mu)^2}{(\Delta x^\mu)^2} + \frac{(i\hbar\partial_\mu - P_\mu)^2}{(\Delta p_\mu)^2}\right] \quad (7.5)$$

respectively with the eigenvalues $(2n^\mu + 1)(\Delta x^\mu)^2$ and $(2n^\mu + 1)(\Delta p_\mu)^2$. We may introduce correlated variables $y^\mu$ as components of a new position quadrivector $\begin{bmatrix} y^0 \\ \vec{y} \end{bmatrix}$ which may be related to $\begin{bmatrix} x^0 \\ \vec{x} \end{bmatrix}$ by a linear transformation $\Lambda$ in the Minkowski space, as in the relation (5.26)

$$\begin{cases} y^\mu = \Lambda_\nu^\mu x^\nu \\ Y^\mu = \Lambda_\nu^\mu X^\nu \\ k_\mu = V_\mu^\nu p_\nu \\ K_\mu = V_\mu^\nu P_\nu \end{cases} \Leftrightarrow \begin{cases} x^\nu = V_\mu^\nu y^\mu \\ X^\nu = V_\mu^\nu Y^\mu \\ p_\nu = \Lambda_\nu^\mu k_\mu \\ P_\nu = \Lambda_\nu^\mu K_\nu \end{cases} \quad (7.6)$$

$V$ is the inverse of $\Lambda$: $V = \Lambda^{-1}$. If we choose $\Lambda$ such as $(det\Lambda)^2 = 1$, then $\Lambda$ is a Lorentz transformation.

We may define for correlated variables the dispersion-codispersion tensors $\mathcal{A} = \mathcal{A}^{\mu\nu}\vec{e}_\mu \otimes \vec{e}_\nu$ and $\mathcal{B} = \mathcal{B}_{\mu\nu}\vec{e}^\mu \otimes \vec{e}^\nu$ such as

$$\mathcal{A}^{\mu\nu} = \sum_{\rho=0}^{3} \frac{\hbar^2 \Lambda_\rho^\mu \Lambda_\rho^\nu}{(2\Delta p_\rho)^2} = \Lambda_\rho^\mu \Lambda_\rho^\nu \Delta x^\rho \Delta x^\rho \quad (7.7)$$

$$\mathcal{B}_{\mu\nu} = \sum_{\rho=0}^{3} \frac{\hbar^2 V_\mu^\rho V_\nu^\rho}{(2\Delta x^\rho)^2} = V_\mu^\rho V_\nu^\rho \Delta p_\rho \Delta p_\rho \quad (7.8)$$

Then we may define the correlated quadrimensional harmonic Gaussian functions $\chi_n^4$ such as

$$\chi_n^4(y^0, \vec{y}, Y^0, \vec{Y}, K_0, \vec{K}, \mathcal{B}) = \varphi_n^4(x^0, \vec{x}, X^0, \vec{X}, P_0, \vec{P}, [\Delta p]) \quad (7.9)$$

These functions are eigenfuctions of the dispersion operators

$$\Sigma_y^{\mu\nu} = \frac{1}{2}(y^\mu - Y^\mu)(y^\nu - Y^\nu)$$
$$+ \frac{2}{\hbar^2}\mathcal{A}^{\mu\alpha}\mathcal{A}^{\nu\beta}(i\hbar\frac{\partial}{\partial y^\alpha} - K_\alpha)(i\hbar\frac{\partial}{\partial y^\alpha} - K_\beta) \quad (7.10)$$

$$\Sigma_{\mu\nu}^k = \frac{1}{2}(i\hbar\frac{\partial}{\partial y^\mu} - K_\mu)\left(i\hbar\frac{\partial}{\partial y^\nu} - K_\nu\right)$$
$$+ \frac{2}{\hbar^2}\mathcal{B}_{\mu\alpha}\mathcal{B}_{\nu\beta}(y^\alpha - Y^\alpha)(y^\beta - Y^\beta) \quad (7.11)$$

We may introduce quadratic mean operators with the dispersion operators. We may particularly define the quadratic mean operators which correspond to the components of the four-momentum covector.

$$\overline{(k_\mu)^2} = \mathbf{\Sigma}^{\mathbf{k}}_{\mu\mu} + (K_\mu)^2 \qquad (7.12)$$

*VII.3 Equation for scalar field*

Let us consider the relation between the components of the four-momentum covector

$$(k_0)^2 - (k_1)^2 - (k_2)^2 - (k_3)^2 = m^2 c^2 \qquad (7.13)$$

If we make the replacement $(k_\mu)^2 \rightarrow \overline{(k_\mu)^2}$ in this relation, we obtain the operatorial relation

$$g^{\mu\nu}[\frac{1}{2}\left(i\hbar\frac{\partial}{\partial y^\mu} - K_\mu\right)\left(i\hbar\frac{\partial}{\partial y^\nu} - K_\nu\right)$$
$$+ \frac{2}{\hbar^2}\mathcal{B}_{\mu\alpha}\mathcal{B}_{\nu\beta}(y^\alpha - Y^\alpha)(y^\beta - Y^\beta)] + M^2 c^2 = \overline{m^2} c^2 \quad (7.14)$$

in which we have introduce a mass quadratic mean operator $\overline{m^2}$ and the mass mean value $M$ defined by the relation

$$g^{\mu\nu} K_\mu K_\nu = M^2 c^2 \qquad (7.15)$$

From the operatorial relation (7.14), we may deduce, as in the case of Klein-Gordon equation, an equation for scalar field in the framework of our approach. This equation is

$$\{g^{\mu\nu}[\frac{1}{2}\left(i\hbar\frac{\partial}{\partial y^\mu} - K_\mu\right)\left(i\hbar\frac{\partial}{\partial y^\nu} - K_\nu\right)$$
$$+ \frac{2}{\hbar^2}\mathcal{B}_{\mu\alpha}\mathcal{B}_{\nu\beta}(y^\alpha - Y^\alpha)(y^\beta - Y^\beta)] - (\overline{m^2} - M^2)c^2\}\phi = 0$$

This results show that our approach may be used in formulation of relativistic field theory. More depthful studies on the physical meaning of the results that we have obtained for the case of scalar field and extension to the case of spinorial and vectorial fields may lead to more interesting results.

## VII. CONCLUSION

The results obtained in section II, III and IV show that properties of harmonic Gaussian functions may be used to introduce phase space representation in quantum mechanics for the case of one dimensional motion. According to the relation (4.1) and (4.2), the basis states introduced for this representation can be considered as eigenstates of dispersion operators.

It was shown in the section V that the results obtained for the case of one dimension may be generalized to multidimensional cases.

The examples of application described in the sections VI and VII show that our approach may be applied both in formulation of nonrelativstic and relativistic theory.

Results thus obtained show that our approach may be considered as a possible method to establish a framework for formulation of quantum theory in phase space.